\documentstyle[preprint,aps]{revtex}
\begin{document}
\draft


\title{ The Spallagenic Production Rates of Lithium, Beryllium and Boron }

\author{ J.P. Kneller$^{1}$ \and J.R. Phillips$^{1}$ \and T.P.Walker$^{1,2}$ } 
\address{ $^1$ Department of Physics, The Ohio State University, Columbus, OH 43210 \\
          $^2$ Department of Astronomy, The Ohio State University, Columbus, OH 43210
         }
\date{\today}

\maketitle

\newcommand{\LiSix}{$^{6} \rm Li$}
\newcommand{\LiSeven}{$^{7} \rm Li$}
\newcommand{\BeNine}{$^{9} \rm Be$}
\newcommand{\BTen}{$^{10} \rm B$}
\newcommand{\BEleven}{$^{11} \rm B$}


\begin{abstract}

We calculate the production rates of \LiSix, \LiSeven, \BeNine, \BTen\ and
\BEleven\ via spallation of Carbon, Nitrogen and Oxygen nuclei by protons
and $\alpha$-particles and by $\alpha-\alpha$ fusion reactions. We include recent
measurements of the cross sections of $\alpha$-$\alpha$ fusion reactions and
find that the computations yield rates of \LiSix\ and \LiSeven\
production that are nearly a factor of two smaller than previously calculated. 
We begin by using the `straight ahead' approximation for
the fragment energy and the `leaky-box' model for product capture in the
Galaxy. In addition we test the straight ahead approximation by
recalculating the production rates using an empirical description of the
fragment energy distribution and find that the results closely match. We
have also calculated the rates for various cosmic ray spectra and find
that the hardest spectra tested decrease the rates with CR CNO by
approximately an order of magnitude relative to our chosen standard. 
Finally we have computed the Population I elemental ratios and the Population II
scaling relations for our standard and find that our computations predict an abundance of
Lithium for a given abundance of Beryllium that is 1/4 smaller than previously derived.

\end{abstract}


\section{Introduction}

The spallation production rates of the isotopes of Lithium, Beryllium and
Boron (LiBeB) are a necessary component in any calculation of the
evolution of these nuclei in the Galaxy. In particular, \LiSix, \BeNine\
and \BTen\ are thought to be produced solely by spallation and thus give
clues to the contribution of spallagenic \LiSeven\ and \BEleven\ to the
total abundance of these nuclei. For example it is the comparison of the
observed solar \BEleven/\BTen\ isotopic ratio of 4.05 ~\cite{AG1989} to
the spallagenic prediction of $\sim$2.5 ~\cite{R1974} that has led many
authors to include the production of \BEleven\ by supernova
into their models of Galactic chemical evolution~\cite{VFea1998}. LiBeB
have also been observed in many metal poor halo
stars~\cite{Mea1997,DLL1992,Dea1997,GLea1998,HTR1999,NLPS1999,SLN1993} and
the data indicates that the abundances of these light nuclei are
linearly (or almost linearly) proportional to the abundance of Iron in
these stars and contrary to the standard model
prediction of a quadratic proportion that would be
expected if the light nuclei were a secondary product~\cite{FO1999}. This
would seem to indicate either
a primary origin by some other mechanism or a production rate dominated by
spallation of CRs enriched in CNO on interstellar p and $\alpha$. It is
the purpose of this paper to compute the spallagenic rates of production
of these light nuclei with inclusion of new $\alpha - \alpha$ cross
section data and to examine the details of these calculations: in
particular, to test the `straight ahead' approximation to the fragment
energy distribution. 
In the last section we use our new production rates and rederive the elemental 
ratios for the Population I and the scaling relations for the Population II environments 
previously derived in Steigman \& Walker~\cite{SW1992}.


\section{The Straight Ahead Calculation}
\label{sec:SAC}

The production rates $J^{PTF}$ for the reaction $\rm P + T \rightarrow F + ...$ are formally calculated from the equation
\begin{equation} 
J^{PTF} = \int dE_{P} \ \frac{d \phi_{P} (E_{P})} {dE_{P}} \ \sigma^{P+T \rightarrow F} (E_{P}) \left\{ \int dE_{F} \ \rho(E_{F}[E_{P}] ) \ S_{F}(E_{F}) \right\}, \label{eq:Rsa} 
\end{equation} 
where P labels the incident CR projectile, T the ISM target and F the
fragment produced. $d\phi_{P}(E_{P})/dE_{P}$ is the interstellar cosmic
ray spectrum, $\sigma^{P+T \rightarrow F} (E_{P})$ the relevant cross section for
the process, $\rho( E_{F}[E_{P}] )$ is the fragment energy distribution
function and $S_{F}(E_{F})$ is a factor that accounts for the successful
trapping of the fragment F in the Galaxy. Here, and in the remainder of
this paper, all energies/momenta are in per nucleon units. At present
there is no accepted standard form for the CR spectrum since the
interstellar cosmic ray spectrum below $\sim \rm 5 \ GeV/nucleon$ cannot
be measured directly~\cite{M1997}. The cosmic ray spectrum as measured by
balloon borne experiments at the top of the atmosphere~\cite{SOSSJSB1991},
by satellites or instruments carried into orbit~\cite{GLMMS1988} or
by interplanetary spacecraft~\cite{CCS1995} has been modulated by the
geomagnetic effects of the sun. Theoretical descriptions of this
modulation use these observations as tests for the models and all require
the interstellar cosmic ray spectrum as input~\cite{M1997,IA1985}. Many
studies of chemical evolution~\cite{FOS1994,RKLR1997}) employ a technique
where the interstellar cosmic ray spectrum is formed from propagating a
\emph{source} spectrum through the ISM.  For the present we choose a CR
spectrum of the form
\begin{equation} 
\frac{d \phi_{P} (E_{P})}{dE_{P}} = \frac{ 1.6 \ E^{1.6}_{0} }{ ( E_{P} + E_{0} )^{2.6} } \ \ \rm{/(MeV/nucleon)/(cm^{2}/s)} \label{eq:dphi_de_WMV} 
\end{equation} 
which was presented by Gloeckler \& Jokipii~\cite{GJ1967} and used by
Walker, Matthews \& Viola~\cite{WMV1985}, Steigman \& Walker~\cite{SW1992}
and Urch \& Gleeson~\cite{UG1973}. The parameter $E_{0}$ is the nucleon
mass. We shall consider modifications of this form in
section~\ref{sec:TDS}.

If the Galaxy trapped every fragment produced by the reaction $\rm P+T
\rightarrow F + ...$ then $S_{F}(E_{F}) = 1$ and the integral over the
fragment energy $E_{F}$ is trivial. This is true in the cases of p and
$\alpha$ as projectiles, however this is not the case when the projectile
is CNO. The fragment energy distribution $\rho( E_{F}[E_{P} ])$ and
S-factor are usually taken to be
\begin{equation} 
\rho(E_{F}[E_{P}]) = \delta( E_{P} - E_{F} ), \label{eq:delta} 
\end{equation}
and
\begin{equation}
\\ S_{F}(E_{F}) = \left\{ \begin{array}{cr} 1                                              & \ \ \ \rm{ P \in \{H,He\} }  \\ 
                                                \exp \left[ -R_{F}(E_{F})/ \Lambda \right] & \ \ \ \rm{ P \in \{C,N,O\} } \\
                              \end{array} \right\}, \label{eq:S} 
\end{equation} 
where $R_{F}(E_{F})$ is the range of F and $\Lambda$ is the stopping power
of the ISM. The $\delta$ function form for $\rho( E_{F}[E_{P} )]$ is known
as the `straight ahead' approximation and the form of $S_{F}(E_{F})$ is
often called the `leaky-box' model~\cite{C1980}. The range
$R_{F}(E_{F})$ is tabulated up to 12 MeV/nucleon for each fragment by
Northcliffe \& Schilling~\cite{NS1970}. Above this energy we use the range
for protons as tabulated by Janni~\cite{J1982} and rescale in exactly the
same manner as Fields, Olive \& Schramm~\cite{FOS1994}. We assume that
$\Lambda$ is independent of energy although Fields, Olive \&
Schramm~\cite{FOS1994} introduced an energy dependent form while Tsao,
Silberberg, Barghouty \& Sihver~\cite{TSBS1995} use a rigidity dependent
$\Lambda$ introduced by Gupta \& Webber~\cite{GW1989}.

The cross sections needed to create LiBeB were tabulated by Read \&
Viola~\cite{RV1984} and we have included the recent measurements of the
$\alpha - \alpha$ fusion reaction cross sections by Mercer et
al~\cite{Mea2000} and the reanalysis by Mercer, Glagola and Austin~\cite{MGA1996}. 
The new measurements show that both these cross sections fall approximately exponentially with projectile energy
and have reduced the uncertainty in these rates due to the necessity of extrapolation to higher energies.  

There are some complications in these computations that are frequently
left unstated in the literature: the cross sections in Read \&
Viola~\cite{RV1984} are for the isobars, in the case of $\rm A=7$ this is
a $\sim$ 50:50 mixture of \LiSeven\ and $^{7}$Be and likewise for $\rm
A=11$ which contains both \BEleven\ and $^{11}$C. These other nuclei will
eventually decay to \LiSeven\ and \BEleven\ respectively and even though
the half-lives are normally short at 53.28 days and 20.3 minutes
respectively~\cite{CRC1999} this cannot occur until they have been trapped
by the Galaxy since the decay mode is inner orbital electron capture.
Thus, for $R_{\rm ^{7}Li}$ and $R_{\rm ^{11}B}$ we average the trapping
factors of these products. In a similar manner the $\rm A=6$ isobaric
cross section includes the production of $^{6}$He but in this case the
decay mode is $\beta$ decay with a half-life of 0.807 s~\cite{CRC1999} and
so any $^{6}$He produced will quickly decay to \LiSix\ and no averaging is
required.
  
In table~\ref{tab:1}: columns a) and b) we present the rates as calculated
by equation (\ref{eq:Rsa}) using the forms for the various terms as listed
in equations (\ref{eq:dphi_de_WMV}) through (\ref{eq:S}) using two
different values of $\Lambda$. An initial examination of the results shows
that fragment loss for the reactions with $\rm P \in \{C,N,O\}$ reduces
the production rate by a factor $\alt 10$ relative to the rate when the
projectile and target are interchanged. While this may appear to render
these reactions unimportant it must be remembered that, at the present
time, the abundances of CNO in the CR are enhanced relative to their ISM
values by approximately equivalent factors. We also find that the new
$\alpha - \alpha$ rates are smaller than those calculated by Walker,
Matthews \& Viola~\cite{WMV1985}: the rate for $\alpha + \alpha
\rightarrow \rm ^{6}Li$ decreases by a factor of 1.52 while the rate for
production of \LiSeven\ decreased by a factor of 1.70. Thus for fixed
projectile fluxes and target densities, the production of Li (that
includes an $\alpha$-$\alpha$\ contribution) relative to that of Be and B
(that does not include $\alpha$-$\alpha$) decreases.


\section{Testing The `Straight Ahead' Approximation} 
\label{sec:test}

In section~\ref{sec:SAC} we used the `straight ahead' approximation to
calculate $J^{PTF}$ and it is this that we now wish to test in the same
manner as Tsao, Silberberg, Barghouty \& Sihver~\cite{TSBS1995}.
The momentum distributions of the isotopes produced by fragmentation of
various projectile and target nuclei are discussed by
Morrissey~\cite{M1989} and Hufner~\cite{H1985}. They are found to be
Gaussian distributed in the rest frame of the spalled nucleus with
narrow dispersions and a small mean momentum in the direction parallel to
the incident projectile. In particular the momentum distributions of the
fragments from $^{12}$C and $^{16}$O projectiles upon various targets
ranging in mass from Be to Pb were measured by Greiner et
al~\cite{Gea1975} who found that their results had no significant
correlation with target mass or beam energy. Goldhaber~\cite{G1974} and
more recently Bauer~\cite{B1995} explain these results in terms of a
nuclear model with minimal correlations between the nucleon momenta and
predict the dependence of the momentum per nucleon distribution
$\rho(\bf{P}_{\it F})$ on fragment mass A$_{F}$ as 
\begin{eqnarray} 
\rho({\bf P}_{\it F} ) & = & \frac{1}{ ( 2 \pi \alpha_{F}^{2} )^{3/2} } \exp \left( \frac{-({\bf P}_{\it F} - <{\bf P}_{\it F}>)^{2} }{ 2 \alpha_{F}^{2} } \right), \label{eq:rhoPf} \\ 
\nonumber \\
\alpha_{F}^{2} & = & \alpha_{0}^{2} \ \frac{ A_{CNO} - A_{F} }{ A_{F}(A_{CNO}-1) },  
\end{eqnarray}
and
\begin{equation}
<{\bf P}_{\it F}> = 8 \ \frac{(A_{CNO}-A_{F}) }{ A_{CNO} } \ \frac{ \gamma + 1 }{ \beta \gamma} \label{eq:meanp} \ \ \ \rm{ MeV/nucleon}, 
\end{equation} 
where $\alpha_0 \sim 100$ MeV, $A_{CNO}$ is
the mass of the nucleus to be spalled while $\beta=v/c$ and
$\gamma=(1-\beta^{2})^{-1/2}$ are those of the projectile. Equation
(\ref{eq:rhoPf}) assumes that the transverse momentum dispersion is equal
to the longitudinal and equation (\ref{eq:meanp}) is a semi-empirical
relation by Morrissey~\cite{M1989}. We also ignore the change in $\alpha$
that occurs for values of E$_{P} \alt 100$ MeV/nucleon indicated by
Stokstad~\cite{S1984}: at such small energies the value of $S_{F}$ is very
close to unity and every fragment isotope is captured by the Galaxy
rendering the decrease in $\alpha$ irrelevant. Thus we can rewrite
equation (\ref{eq:Rsa}) for reactions involving CNO that includes this
better approximation to the product fragmentation energy distribution: 

\begin{eqnarray} 
J^{PTF} & = & \int dE_{P} \frac{1.6 E_{0}^{1.6}}{(E_{P} + E_{0})^{2.6}} \ \sigma^{P+T \rightarrow F} (E_{P}) \nonumber \\ 
        &   & \left\{ \int \frac{d^3 \bf{P}'_{\it F} }{ (2 \pi \alpha_{F}^{2})^{3/2} } \exp \left( \frac{- (\bf{P}'_{\it F} -
      <\bf{P}'_{\it F}>)^{2} }{ 2 \alpha_{F}^2} \right) \ \exp \left(\frac{-R_{F}(E_{F}) }{\Lambda} \right) \right\}. \label{eq:Rp}
\end{eqnarray} 
The dependence of $E_{P}$ upon $E_{F}$ is more subtle
than before and enters into the integral over momentum per nucleon
$\bf{P}'_{\it F}$ since this is evaluated in the CNO rest frame.

In table~\ref{tab:1}: columns c) and d) we present the rates as calculated
by equation (\ref{eq:Rp}) again for $\Lambda = \rm 5 \ g/cm^{2}$ and $\rm
10 \ g/cm^{2}$. A glance at the results shows that indeed the `straight
ahead' approximation accurately (within numerical error) predicts the
production rate of those reactions with $\rm P \in \{p,\alpha\}$. This
result was to be expected: the spread in fragment momenta $\alpha_{F}$ in
equation (\ref{eq:Rp}) is only of order $\sim \rm 100 \ MeV$ and the
central value corresponds to a stopping range much less than $\Lambda$ so
all fragments are trapped. For the reactions with $\rm P \in \{C,N,O\}$
there appears to be a slight increase in the production rates relative to
the `straight ahead' calculations for the smaller value of $\Lambda$ and a
slight decrease for the larger value of $\Lambda$ but the differences are
small and do not warrant rejection of the approximation. Again, the
approximation's success has a simple explanation: the fragments that
escape must have large momenta ($\agt \rm 500 \ MeV$) and the small values
of $\alpha_{F}$ mean that $\bf{P}_{\it F} \sim \bf{P}_{\it P}, \it E_{F}
\sim E_{P}$.


\section{Changing The CR Spectrum}
\label{sec:TDS}

The CR spectrum (\ref{eq:dphi_de_WMV}) used in sections \ref{sec:SAC} and
\ref{sec:test} is by no means the only one that has been used. As
previously stated, solar modulation significantly depresses the low energy
($\alt \rm 5 \ GeV$) region of the CR spectrum and the extent of this
alteration can be mitigated by changing the unmodulated spectrum to yield
similar observations. Yet it is exactly this region that dominates the
integrands in equation (\ref{eq:Rp}) since it corresponds to where
$S_{F}(E_{F})$ is $\approx 1$. To address this uncertainty we also
calculate the production rates $J^{PTF}$ with interstellar spectra that
differ from equation (\ref{eq:dphi_de_WMV}). More specifically, we rewrite
equation (\ref{eq:dphi_de_WMV}) as 
\begin{equation} 
\frac{d \phi_{P} (E_{P})}{dE} = \frac{ E_{0}^{\nu-\mu-1} }{ B(\mu+1,\nu-\mu-1) } \ \frac{E^{\mu}_{P}}{ ( E_{P} + E_{0} )^{\nu} } \ \ \rm /(MeV/nucleon)/(cm^{2}/s), \label{eq:dphi_de_e_numu} 
\end{equation} 
where $B(\mu+1,\nu-\mu-1)$ is the Beta function. This functional form is
sufficiently flexible to mimic virtually any spectrum desired but, to avoid
a plethora of results, we restrict ourselves to considering values of
$\mu$ and $\nu$ such that $\nu-\mu = 2.6, 0 \leq \mu \leq 1$ and keep
$E_{0}$ at it's previous value. This family of spectra asymptotically
approaches the form of equation (\ref{eq:dphi_de_WMV}) at high energies but
`push' some of their low energy component to higher values relative to the
$\mu=0$,$\nu=2.6$ standard. Consequently it should be expected that the
rates for CR CNO on ISM H,He would be reduced due to a decrease in the
fraction of captured fragments. The rates with $\rm P \in \{p,\alpha\}$
will be affected but only to a slight degree since all fragments are still
captured but the cross sections are not constants. The effect of the
reduction of the spectrum at lower energies on the $\alpha + \alpha$
fusion production rates of \LiSix\ and \LiSeven\ should be more pronounced
because the cross sections for both of these processes are predominant at
lower energies. The results of using equation (\ref{eq:dphi_de_e_numu})
are listed in table \ref{tab:2} for the values $\mu \in
\{0,0.2,0.4,0.6,0.8,1.0\}$ and $\Lambda$ fixed at 5 g/cm$^{2}$. As
predicted the rates for $\rm P \in \{p,\alpha\}$ are almost unaltered but
the rates for the reactions with $\rm P \in \{C,N,O\}$ and $\alpha +
\alpha \rightarrow {^{6} \rm{Li},^{7}\rm{Li} }$ are significantly affected
by the change in the CR spectra, indeed the rates for $\mu=1,\nu=3.6$ are
approximately an order of magnitude smaller than the rates for
$\mu=0,\nu=2.6$. If these hard spectra are indeed indicative of the true
interstellar cosmic ray spectrum then even the present CR CNO enrichment
mentioned previously cannot bring the `inverse' reactions into competition
with their `forward' brethren and the production of the \BeNine\ and
\BTen\ in the Galaxy by this process has always been dominated by
reactions with p and $\alpha$ as the projectiles.


\section{Population I Elemental Ratios and Population II Scaling Relations}

Finally we have recomputed the scaling relations and elemental ratios previously presented in Steigman\&Walker~\cite{SW1992} for our standard case of $\Lambda=$5 g/cm$^{2}$, $\mu=0$,$\nu=2.6$. 
The approximate nucleosynthetic yields can be written as
\begin{equation}
y_{F} \approx \sum_{P} \sum_{T} \alpha_{P} \ y_{T} \ J^{PTF} \Delta t = 1.18{\rm x}10^{-12} \ R_{F} \ \Delta t_{Gyr}.
\end{equation}
The uncertainty in the time dependence of the CR spectrum, the S-factor and the abundances of the targets and CRs can be largely removed by considering the ratios of the elements rather than individual yields. The Population I ISM is taken to be \\
\hfill
\parbox{8cm}{ \begin{eqnarray*}
              y_{H}  & = & 1,   \\
              y_{He} & = & 0.1, \\
              y_{C}  & = & 4.2 \ {\rm x} \ 10^{-4}, \\
              y_{N}  & = & 8.7 \ {\rm x} \ 10^{-5}, \\ 
              y_{O}  & = & 6.9 \ {\rm x} \ 10^{-4},
              \end{eqnarray*}
             }
\hfill
\parbox{8cm}{ \begin{eqnarray*}
              \alpha_{p}      & = & 8.05,            \\
              \alpha_{\alpha} & = & \alpha_{p}/10,   \\
              \alpha_{C}      & = & \alpha_{p}/310,  \\
              \alpha_{N}      & = & \alpha_{p}/5120, \\ 
              \alpha_{O}      & = & \alpha_{p}/240,
              \end{eqnarray*}
             } \\
and we compute 
\begin{eqnarray}
R_{7}/R_{6} & = & 1.49,                        \\
R_{7}/R_{9} = 6.22, & & R_{6+7}/R_{9} = 10.39, \\
R_{6+7}/R_{10+11} & = & 0.62,                  \\
R_{11}/R_{10} & = & 2.48, 
\end{eqnarray}
We find that in comparison with S\&W that the our new calculations for the production rates have increased the $7/6$ ratio slightly and reduced all the remainder except for the $11/10$ ratio which has remained unaltered. The inclusion of the `reverse' reactions and the lower \LiSix\ and \LiSeven\ yields have both decreased the denominators and increased the numerators.

In the early Galaxy we adopt the same scaling relations as S\&W, namely \\
\hfill
\parbox{8cm}{ \begin{eqnarray*}
              y^{II}_{H}  & = & 1,    \\
              y^{II}_{He} & = & 0.08, \\
              y^{II}_{C}  & = & y^{I}_{C} \ {\rm x} \ 10^{[Fe/H]}, \\
              y^{II}_{N}  & = & y^{I}_{N} \ {\rm x} \ 10^{[Fe/H]}, \\ 
              y^{II}_{O}  & = & y^{I}_{O} \ {\rm x} \ 10^{[Fe/H]+1/2}, 
              \end{eqnarray*}
             }
\hfill
\parbox{8cm}{ \begin{eqnarray*}
              \alpha^{II}_{p}      & = & \alpha^{I}_{p},                            \\
              \alpha^{II}_{\alpha} & = & \alpha^{II}_{p} \ {\rm x} \ 0.08,          \\
              \alpha^{II}_{C}      & = & \alpha^{I}_{C} \ {\rm x} \ 10^{[Fe/H]},    \\
              \alpha^{II}_{N}      & = & \alpha^{I}_{N} \ {\rm x} \ 10^{[Fe/H]},    \\ 
              \alpha^{II}_{O}      & = & \alpha^{I}_{O} \ {\rm x} \ 10^{[Fe/H]+1/2},
              \end{eqnarray*}
             } \\
and find 
\begin{eqnarray}
R_{6} = 1.27 ( 1+16.94 \ {\rm x} \ 10^{[Fe/H]}) & & R_{7} = 2.09 ( 1+13.95 \ {\rm x} \ 10^{[Fe/H]}) \\
& R_{9} = 5.97 \ {\rm x} \ 10^{[Fe/H]} & \\
R_{10} = 25.98 \ {\rm x} \ 10^{[Fe/H]} & & R_{11} = 60.74 \ {\rm x} \ 10^{[Fe/H]}                                
\end{eqnarray}
Once again, in comparison with S\&W: the contribution of the $\alpha-\alpha$ reactions to the production of \LiSix\ and \LiSeven\ have almost halved the constant in $R_{6}$ and $R_{7}$ and the CNO terms have doubled. The \BeNine,\BTen\ and \BEleven\ yields are approximately doubled too. While the new yields do not alter the basic conclusions of S\&W they do make it weaker and the amount of \LiSix\ and \LiSeven\ that would be inferred by observation of the amount of \BeNine\ in the oldest stars would be reduced. For example: at a metalicity of $[Fe/H]=-3$ the ratio of Lithium to Beryllium is now only $\sim 550$ compared to a value of $\sim 2200$ found in Steigman \& Walker, a factor of 4 difference. We also obtain a better agreement with the \LiSix\ to \BeNine\ ratios observed in metal-poor halo stars: for the star HD84937 Hobbs, Thorburn \& Rebull~\cite{HTR1999} find the ratio to be $73 \pm 18$ whereas the prediction is $37.3$: for BD$+$26$^{\rm \circ}$3578 the observed ratio is $22 \pm 13$ and the prediction is $57.0$.
  

\section{Conclusions}

We have calculated the production rates of \LiSix, \LiSeven, \BeNine,
\BTen\ and \BEleven\ via spallation of Carbon, Nitrogen and Oxygen nuclei
by protons and $\alpha$-particles and the production of \LiSix\ and
\LiSeven\ by $\alpha-\alpha$ fusion reactions. We have found that the new
$\alpha - \alpha$ fusion cross section data produces smaller production
rates than previously computed by factors of $\sim$ 1.5 and 1.7 for
\LiSix\ and \LiSeven\ respectively. By employing a better description of
the fragment energy as a function of the projectile's energy we relaxed
the `straight ahead' approximation and found that the rates were only
slightly affected. We also computed the production rates with increasingly
harder spectra and found that they decreased by up to an order of
magnitude compared to our reference values. We conclude that, if these
spectra represent the true interstellar spectrum, the \BeNine\ and \BTen\
production in the Galaxy is always dominated by the production from the
`forward' reactions of CR p/$\alpha$ upon CNO in the ISM.


\begin{table}									
\begin{tabular}{||l|c|c|c|c||}									
rate (P + T $\rightarrow$ F) & a) & b) & c) & d) \\ \hline									

H + C $\rightarrow$ \LiSix  	&	1.21E-26	&	1.21E-26	&	1.21E-26	&	1.21E-26	\\
C + H $\rightarrow$ \LiSix  	&	1.57E-27	&	2.37E-27	&	1.68E-27	&	2.36E-27	\\
H + C $\rightarrow$ \LiSeven   	&	2.15E-26	&	2.15E-26	&	2.15E-26	&	2.14E-26	\\
C + H $\rightarrow$ \LiSeven   	&	3.80E-27	&	5.35E-27	&	3.98E-27	&	5.30E-27	\\
H + C $\rightarrow$ \BeNine  	&	4.09E-27	&	4.09E-27	&	4.10E-27	&	4.10E-27	\\
C + H $\rightarrow$ \BeNine  	&	3.97E-28	&	6.19E-28	&	4.10E-28	&	6.22E-28	\\
H + C $\rightarrow$ \BTen   	&	2.28E-26	&	2.28E-26	&	2.28E-26	&	2.28E-26	\\
C + H $\rightarrow$ \BTen   	&	4.22E-27	&	6.05E-27	&	4.29E-27	&	6.07E-27	\\
H + C $\rightarrow$ \BEleven   	&	5.73E-26	&	5.73E-26	&	5.73E-26	&	5.73E-26	\\
C + H $\rightarrow$ \BEleven   	&	1.43E-26	&	1.90E-26	&	1.44E-26	&	1.90E-26	\\ \hline
									
H + N $\rightarrow$ \LiSix   	&	1.82E-26	&	1.82E-26	&	1.82E-26	&	1.82E-26	\\
N + H $\rightarrow$ \LiSix   	&	2.61E-27	&	3.79E-27	&	2.81E-27	&	3.74E-27	\\
H + N $\rightarrow$ \LiSeven   	&	1.05E-26	&	1.05E-26	&	1.05E-26	&	1.05E-26	\\
N + H $\rightarrow$ \LiSeven   	&	2.24E-27	&	2.95E-27	&	2.34E-27	&	2.94E-27	\\
H + N $\rightarrow$ \BeNine  	&	4.67E-27	&	4.67E-27	&	4.66E-27	&	4.66E-27	\\
N + H $\rightarrow$ \BeNine  	&	8.24E-28	&	1.17E-27	&	8.57E-28	&	1.16E-27	\\
H + N $\rightarrow$ \BTen   	&	1.16E-26	&	1.16E-26	&	1.16E-26	&	1.16E-26	\\
N + H $\rightarrow$ \BTen   	&	3.58E-27	&	4.40E-27	&	3.65E-27	&	4.39E-27	\\
H + N $\rightarrow$ \BEleven   	&	2.41E-26	&	2.41E-26	&	2.41E-26	&	2.41E-26	\\
N + H $\rightarrow$ \BEleven   	&	6.93E-27	&	8.76E-27	&	7.02E-27	&	8.76E-27	\\ \hline
									
H + O $\rightarrow$ \LiSix   	&	1.29E-26	&	1.29E-26	&	1.29E-26	&	1.29E-26	\\
O + H $\rightarrow$ \LiSix   	&	1.99E-27	&	2.85E-27	&	2.12E-27	&	2.83E-27	\\
H + O $\rightarrow$ \LiSeven  	&	2.06E-26	&	2.06E-26	&	2.06E-26	&	2.06E-26	\\
O + H $\rightarrow$ \LiSeven    	&	3.50E-27	&	5.02E-27	&	3.73E-27	&	4.95E-27	\\
H + O $\rightarrow$ \BeNine    	&	4.12E-27	&	4.12E-27	&	4.12E-27	&	4.13E-27	\\
O + H $\rightarrow$ \BeNine  	&	7.11E-28	&	1.03E-27	&	7.48E-28	&	1.01E-27	\\
H + O $\rightarrow$ \BTen  	&	1.49E-26	&	1.49E-26	&	1.49E-26	&	1.49E-26	\\
O + H $\rightarrow$ \BTen  	&	3.11E-27	&	4.31E-27	&	3.22E-27	&	4.26E-27	\\
H + O $\rightarrow$ \BEleven  	&	2.79E-26	&	2.79E-26	&	2.79E-26	&	2.79E-26	\\
O + H $\rightarrow$ \BEleven  	&	7.70E-27	&	1.00E-26	&	7.88E-27	&	1.00E-26	\\ \hline

He + C $\rightarrow$ \LiSix   	&	4.36E-26	&	4.36E-26	&	4.36E-26	&	4.36E-26	\\
C + He $\rightarrow$ \LiSix   	&	9.03E-27	&	1.21E-26	&	9.54E-27	&	1.19E-26	\\
He + C $\rightarrow$ \LiSeven  	&	5.97E-26	&	5.97E-26	&	5.97E-26	&	5.98E-26	\\
C + He $\rightarrow$ \LiSeven  	&	1.33E-26	&	1.77E-26	&	1.38E-26	&	1.76E-26	\\
He + C $\rightarrow$ \BeNine  	&	1.19E-26	&	1.19E-26	&	1.20E-26	&	1.19E-26	\\
C + He $\rightarrow$ \BeNine  	&	3.45E-27	&	4.36E-27	&	3.53E-27	&	4.38E-27	\\
He + C $\rightarrow$ \BTen  	&	4.94E-26	&	4.94E-26	&	4.94E-26	&	4.94E-26	\\
C + He $\rightarrow$ \BTen  	&	1.28E-26	&	1.67E-26	&	1.30E-26	&	1.66E-26	\\
He + C $\rightarrow$ \BEleven  	&	9.32E-26	&	9.32E-26	&	9.33E-26	&	9.33E-26	\\
C + He $\rightarrow$ \BEleven  	&	2.53E-26	&	3.28E-26	&	2.55E-26	&	3.28E-26	\\ \hline
									
He + N $\rightarrow$ \LiSix   	&	1.55E-26	&	1.55E-26	&	1.55E-26	&	1.55E-26	\\
N + He $\rightarrow$ \LiSix   	&	3.03E-27	&	3.98E-27	&	3.19E-27	&	3.95E-27	\\
He + N $\rightarrow$ \LiSeven  	&	2.18E-26	&	2.18E-26	&	2.18E-26	&	2.18E-26	\\
N + He $\rightarrow$ \LiSeven  	&	3.97E-27	&	5.49E-27	&	4.17E-27	&	5.44E-27	\\
He + N $\rightarrow$ \BeNine  	&	7.40E-27	&	7.40E-27	&	7.38E-27	&	7.40E-27	\\
N + He $\rightarrow$ \BeNine  	&	1.54E-27	&	2.06E-27	&	1.59E-27	&	2.04E-27	\\
He + N $\rightarrow$ \BTen  	&	3.57E-26	&	3.57E-26	&	3.57E-26	&	3.57E-26	\\
N + He $\rightarrow$ \BTen  	&	8.43E-27	&	1.11E-26	&	8.62E-27	&	1.11E-26	\\
He + N $\rightarrow$ \BEleven  	&	7.08E-26	&	7.08E-26	&	7.09E-26	&	7.09E-26	\\
N + He $\rightarrow$ \BEleven  	&	1.60E-26	&	2.18E-26	&	1.63E-26	&	2.18E-26	\\ \hline
									
He + O $\rightarrow$ \LiSix   	&	1.28E-26	&	1.28E-26	&	1.28E-26	&	1.28E-26	\\
O + He $\rightarrow$ \LiSix  	&	2.17E-27	&	2.98E-27	&	2.32E-27	&	2.93E-27	\\
He + O $\rightarrow$ \LiSeven  	&	1.79E-26	&	1.79E-26	&	1.79E-26	&	1.78E-26	\\
O + He $\rightarrow$ \LiSeven  	&	3.21E-27	&	4.46E-27	&	3.39E-27	&	4.43E-27	\\
He + O $\rightarrow$ \BeNine  	&	6.88E-27	&	6.88E-27	&	6.87E-27	&	6.86E-27	\\
O + He $\rightarrow$ \BeNine  	&	1.27E-27	&	1.76E-27	&	1.32E-27	&	1.75E-27	\\
He + O $\rightarrow$ \BTen  	&	2.16E-26	&	2.16E-26	&	2.16E-26	&	2.16E-26	\\
O + He $\rightarrow$ \BTen  	&	4.54E-27	&	6.22E-27	&	4.69E-27	&	6.20E-27	\\
He + O $\rightarrow$ \BEleven  	&	3.84E-26	&	3.84E-26	&	3.84E-26	&	3.84E-26	\\
O + He $\rightarrow$ \BEleven  	&	8.95E-27	&	1.21E-26	&	9.16E-27	&	1.20E-26	\\ \hline
									
He + He $\rightarrow$ \LiSix   	&	8.58E-28	&	8.58E-28	&		&		\\
He + He $\rightarrow$ \LiSeven  	&	1.41E-27	&	1.41E-27	&		&		\\
\end{tabular}									

\caption{ The reaction rates $J^{PTF}$ in units of /s. There are 4 cases listed: \\									
          a) $\Lambda = 5 \rm \ g/cm^2$: straight-ahead approximation \\									
          b) $\Lambda = 10 \rm \ g/cm^2$: straight-ahead approximation \\									
          c) $\Lambda = 5 \rm \ g/cm^2$: momentum integral equation \\									
          d) $\Lambda = 10 \rm \ g/cm^2$: momentum integral equation \\									
         }									
\label{tab:1}
\end{table}

\begin{table}													
\begin{tabular}{||l|c|c|c|c|c|c||}													
rate (P + T $\rightarrow$ F) & a) &  b) & c) & d) & e) & f) \\  \hline													

H + C $\rightarrow$ \LiSix  	&	1.21E-26	&	1.21E-26	&	1.22E-26	&	1.22E-26	&	1.22E-26	&	1.22E-26	\\
C + H $\rightarrow$ \LiSix  	&	1.57E-27	&	1.21E-27	&	9.20E-28	&	6.86E-28	&	5.05E-28	&	3.70E-28	\\
H + C $\rightarrow$ \LiSeven   	&	2.15E-26	&	2.12E-26	&	2.12E-26	&	2.11E-26	&	2.11E-26	&	2.10E-26	\\
C + H $\rightarrow$ \LiSeven   	&	3.80E-27	&	2.93E-27	&	2.21E-27	&	1.65E-27	&	1.22E-27	&	8.94E-28	\\
H + C $\rightarrow$ \BeNine  	&	4.09E-27	&	4.25E-27	&	4.45E-27	&	4.61E-27	&	4.75E-27	&	4.86E-27	\\
C + H $\rightarrow$ \BeNine  	&	3.97E-28	&	3.19E-28	&	2.50E-28	&	1.93E-28	&	1.48E-28	&	1.12E-28	\\
H + C $\rightarrow$ \BTen   	&	2.28E-26	&	2.27E-26	&	2.29E-26	&	2.29E-26	&	2.29E-26	&	2.29E-26	\\
C + H $\rightarrow$ \BTen   	&	4.22E-27	&	3.38E-27	&	2.64E-27	&	2.05E-27	&	1.57E-27	&	1.19E-27	\\
H + C $\rightarrow$ \BEleven   	&	5.73E-26	&	5.59E-26	&	5.56E-26	&	5.51E-26	&	5.48E-26	&	5.45E-26	\\
C + H $\rightarrow$ \BEleven   	&	1.43E-26	&	1.11E-26	&	8.51E-27	&	6.45E-27	&	4.87E-27	&	3.67E-27	\\ \hline
													
H + N $\rightarrow$ \LiSix   	&	1.82E-26	&	1.80E-26	&	1.81E-26	&	1.81E-26	&	1.81E-26	&	1.81E-26	\\
N + H $\rightarrow$ \LiSix   	&	2.61E-27	&	1.97E-27	&	1.46E-27	&	1.07E-27	&	7.77E-28	&	5.62E-28	\\
H + N $\rightarrow$ \LiSeven   	&	1.05E-26	&	1.01E-26	&	1.01E-26	&	9.97E-27	&	9.92E-27	&	9.88E-27	\\
N + H $\rightarrow$ \LiSeven   	&	2.24E-27	&	1.61E-27	&	1.15E-27	&	8.19E-28	&	5.84E-28	&	4.18E-28	\\
H + N $\rightarrow$ \BeNine  	&	4.67E-27	&	4.62E-27	&	4.65E-27	&	4.64E-27	&	4.64E-27	&	4.63E-27	\\
N + H $\rightarrow$ \BeNine  	&	8.24E-28	&	6.35E-28	&	4.80E-28	&	3.59E-28	&	2.66E-28	&	1.97E-28	\\
H + N $\rightarrow$ \BTen   	&	1.16E-26	&	1.08E-26	&	1.05E-26	&	1.02E-26	&	1.00E-26	&	9.97E-27	\\
N + H $\rightarrow$ \BTen   	&	3.58E-27	&	2.55E-27	&	1.80E-27	&	1.27E-27	&	9.00E-28	&	6.42E-28	\\
H + N $\rightarrow$ \BEleven   	&	2.41E-26	&	2.28E-26	&	2.23E-26	&	2.20E-26	&	2.18E-26	&	2.17E-26	\\
N + H $\rightarrow$ \BEleven   	&	6.93E-27	&	4.88E-27	&	3.48E-27	&	2.53E-27	&	1.86E-27	&	1.38E-27	\\ \hline
													
H + O $\rightarrow$ \LiSix   	&	1.29E-26	&	1.27E-26	&	1.28E-26	&	1.27E-26	&	1.26E-26	&	1.26E-26	\\
O + H $\rightarrow$ \LiSix   	&	1.99E-27	&	1.51E-27	&	1.14E-27	&	8.46E-28	&	6.19E-28	&	4.49E-28	\\
H + O $\rightarrow$ \LiSeven  	&	2.06E-26	&	2.04E-26	&	2.05E-26	&	2.05E-26	&	2.04E-26	&	2.03E-26	\\
O + H $\rightarrow$ \LiSeven    	&	3.50E-27	&	2.75E-27	&	2.10E-27	&	1.59E-27	&	1.18E-27	&	8.73E-28	\\
H + O $\rightarrow$ \BeNine    	&	4.12E-27	&	4.09E-27	&	4.11E-27	&	4.10E-27	&	4.09E-27	&	4.08E-27	\\
O + H $\rightarrow$ \BeNine  	&	7.11E-28	&	5.65E-28	&	4.38E-28	&	3.34E-28	&	2.52E-28	&	1.88E-28	\\
H + O $\rightarrow$ \BTen  	&	1.49E-26	&	1.47E-26	&	1.47E-26	&	1.46E-26	&	1.46E-26	&	1.45E-26	\\
O + H $\rightarrow$ \BTen  	&	3.11E-27	&	2.47E-27	&	1.92E-27	&	1.47E-27	&	1.12E-27	&	8.43E-28	\\
H + O $\rightarrow$ \BEleven  	&	2.79E-26	&	2.70E-26	&	2.66E-26	&	2.62E-26	&	2.59E-26	&	2.56E-26	\\
O + H $\rightarrow$ \BEleven  	&	7.70E-27	&	6.04E-27	&	4.64E-27	&	3.51E-27	&	2.64E-27	&	1.97E-27	\\ \hline

He + C $\rightarrow$ \LiSix   	&	4.36E-26	&	4.18E-26	&	4.11E-26	&	4.04E-26	&	3.98E-26	&	3.94E-26	\\
C + He $\rightarrow$ \LiSix   	&	9.03E-27	&	6.61E-27	&	4.75E-27	&	3.38E-27	&	2.40E-27	&	1.69E-27	\\
He + C $\rightarrow$ \LiSeven  	&	5.97E-26	&	5.76E-26	&	5.67E-26	&	5.59E-26	&	5.54E-26	&	5.49E-26	\\
C + He $\rightarrow$ \LiSeven  	&	1.33E-26	&	9.86E-27	&	7.20E-27	&	5.22E-27	&	3.77E-27	&	2.71E-27	\\
He + C $\rightarrow$ \BeNine  	&	1.19E-26	&	1.12E-26	&	1.08E-26	&	1.04E-26	&	1.02E-26	&	1.00E-26	\\
C + He $\rightarrow$ \BeNine  	&	3.45E-27	&	2.52E-27	&	1.82E-27	&	1.30E-27	&	9.28E-28	&	6.62E-28	\\
He + C $\rightarrow$ \BTen  	&	4.94E-26	&	4.74E-26	&	4.67E-26	&	4.60E-26	&	4.55E-26	&	4.51E-26	\\
C + He $\rightarrow$ \BTen  	&	1.28E-26	&	9.62E-27	&	7.12E-27	&	5.26E-27	&	3.88E-27	&	2.86E-27	\\
He + C $\rightarrow$ \BEleven  	&	9.32E-26	&	8.97E-26	&	8.84E-26	&	8.74E-26	&	8.66E-26	&	8.60E-26	\\
C + He $\rightarrow$ \BEleven  	&	2.53E-26	&	1.90E-26	&	1.42E-26	&	1.07E-26	&	7.98E-27	&	5.98E-27	\\ \hline
													
He + N $\rightarrow$ \LiSix   	&	1.55E-26	&	1.49E-26	&	1.47E-26	&	1.46E-26	&	1.45E-26	&	1.45E-26	\\
N + He $\rightarrow$ \LiSix   	&	3.03E-27	&	2.07E-27	&	1.41E-27	&	9.77E-28	&	6.80E-28	&	4.76E-28	\\
He + N $\rightarrow$ \LiSeven  	&	2.18E-26	&	2.13E-26	&	2.14E-26	&	2.13E-26	&	2.12E-26	&	2.12E-26	\\
N + He $\rightarrow$ \LiSeven  	&	3.97E-27	&	2.94E-27	&	2.16E-27	&	1.58E-27	&	1.15E-27	&	8.42E-28	\\
He + N $\rightarrow$ \BeNine  	&	7.40E-27	&	7.18E-27	&	7.14E-27	&	7.10E-27	&	7.07E-27	&	7.05E-27	\\
N + He $\rightarrow$ \BeNine  	&	1.54E-27	&	1.11E-27	&	8.05E-28	&	5.81E-28	&	4.21E-28	&	3.06E-28	\\
He + N $\rightarrow$ \BTen  	&	3.57E-26	&	3.45E-26	&	3.42E-26	&	3.40E-26	&	3.39E-26	&	3.38E-26	\\
N + He $\rightarrow$ \BTen  	&	8.43E-27	&	6.13E-27	&	4.47E-27	&	3.29E-27	&	2.43E-27	&	1.81E-27	\\
He + N $\rightarrow$ \BEleven  	&	7.08E-26	&	6.97E-26	&	6.98E-26	&	6.97E-26	&	6.96E-26	&	6.95E-26	\\
N + He $\rightarrow$ \BEleven  	&	1.60E-26	&	1.24E-26	&	9.52E-27	&	7.28E-27	&	5.55E-27	&	4.23E-27	\\ \hline
													
He + O $\rightarrow$ \LiSix   	&	1.28E-26	&	1.25E-26	&	1.25E-26	&	1.24E-26	&	1.24E-26	&	1.24E-26	\\
O + He $\rightarrow$ \LiSix  	&	2.17E-27	&	1.56E-27	&	1.11E-27	&	7.89E-28	&	5.60E-28	&	3.98E-28	\\
He + O $\rightarrow$ \LiSeven  	&	1.79E-26	&	1.76E-26	&	1.76E-26	&	1.75E-26	&	1.75E-26	&	1.75E-26	\\
O + He $\rightarrow$ \LiSeven  	&	3.21E-27	&	2.39E-27	&	1.76E-27	&	1.29E-27	&	9.47E-28	&	6.93E-28	\\
He + O $\rightarrow$ \BeNine  	&	6.88E-27	&	6.77E-27	&	6.78E-27	&	6.77E-27	&	6.77E-27	&	6.76E-27	\\
O + He $\rightarrow$ \BeNine  	&	1.27E-27	&	9.54E-28	&	7.09E-28	&	5.25E-28	&	3.87E-28	&	2.85E-28	\\
He + O $\rightarrow$ \BTen  	&	2.16E-26	&	2.12E-26	&	2.13E-26	&	2.12E-26	&	2.12E-26	&	2.11E-26	\\
O + He $\rightarrow$ \BTen  	&	4.54E-27	&	3.47E-27	&	2.62E-27	&	1.98E-27	&	1.49E-27	&	1.12E-27	\\
He + O $\rightarrow$ \BEleven  	&	3.84E-26	&	3.76E-26	&	3.75E-26	&	3.74E-26	&	3.73E-26	&	3.73E-26	\\
O + He $\rightarrow$ \BEleven  	&	8.95E-27	&	6.86E-27	&	5.22E-27	&	3.96E-27	&	3.01E-27	&	2.29E-27	\\ \hline
													
He + He $\rightarrow$ \LiSix   	&	8.58E-28	&	5.32E-28	&	3.26E-28	&	1.99E-28	&	1.23E-28	&	7.77E-29	\\
He + He $\rightarrow$ \LiSeven  	&	1.41E-27	&	7.99E-28	&	4.43E-28	&	2.42E-28	&	1.32E-28	&	7.28E-29	\\
\end{tabular}													

\caption{ The reaction rates $J^{PTF}$ in units of /s. The 6 cases listed: \\ 													
          a) $\nu=0.0$; $\mu=2.6$ (standard) \\													
          b) $\nu=0.2$; $\mu=2.8$ \\													
          c) $\nu=0.4$; $\mu=3.0$ \\													
          d) $\nu=0.6$; $\mu=3.2$ \\													
          e) $\nu=0.8$; $\mu=3.4$ \\													
          f) $\nu=1.0$; $\mu=3.6$ \\	
	 }
\label{tab:2}	
\end{table}


\end{document}